\documentclass{aa}
\usepackage{txfonts}
\usepackage{natbib, graphicx}

\begin{document}
\title{Color Dependence in the Spatial Distribution of Satellite Galaxies}
\author{Jacqueline Chen \inst{1}}
\institute{Argelander-Institut f\"{u}r Astronomie, Universit\"{a}t Bonn, 
Auf dem H\"{u}gel 71,
D-53121 Bonn, Germany; 
       {\tt jchen@astro.uni-bonn.de}}

\begin{abstract}{}
{We explore the color dependence of the radial profile of satellite galaxies
around isolated parent galaxies.}
{Samples of potential satellites selected from
large galaxy redshift surveys are significantly contaminated by
interlopers -- objects not bound to the parent galaxy.  We use the Sloan Digital Sky Survey to estimate the interloper fraction in samples of candidate satellite galaxies.}{We show that samples of red and blue
satellites have different interloper populations: a larger fraction of
blue galaxies are likely to be
interlopers compared to red galaxies.  Both with and without interloper
subtraction, the radial profile of blue satellites is significantly
shallower than that of red satellites.  In addition, while red and blue
primaries have
different interloper fractions, the slope of the corrected radial profiles are
consistent after interloper correction.  We discuss the implications of these
results for galaxy formation models.}
{}
\end{abstract}
\maketitle

\section{Introduction}

In cold dark matter (CDM) models of the universe, large numbers of dark matter (DM) subhalos lie within the virial radius of larger dark matter halos.  In some halos and subhalos, baryonic material has cooled and formed stars, resulting in a central galaxy and satellite galaxies.  The spatial distribution of satellite galaxies in galaxy-sized halos, then, reflects the evolution of satellite galaxies and the mass accretion history of their parent halos.  For example, dark matter simulations suggest that the angular distribution of subhalos follows the shape of the DM halo, which is indicative of infall of subhalos along filaments \citep[e.g.,][]{zentner_etal05,libeskind_etal05}, while observations suggest that satellites lie along the major axis of the light distribution for early-type galaxies \citep[e.g.,][]{sales_lambas04,brainerd05,yang_etal06,azzaro_etal07}.  

In dark matter simulations, the radial distribution of subhalos is biased with respect to the density profile of DM halos;  at small separations from the center of the halo (within $\sim 20-50$\% of the virial radius), the distribution of DM subhalos has a lower concentration, but it follows the dark matter density profile at larger radii \citep{ghigna_etal98,colin_etal99,ghigna_etal00, springel_etal01,delucia_etal04,diemand_etal04,gao_etal04,nagai_kravtsov05,maccio_etal06}.   Studies that include baryons, star formation and cooling, however, show that the distribution of galaxies associated with subhalos has a steeper inner profile than the subhalo distribution, both at cluster and at galaxy scales \citep{nagai_kravtsov05,maccio_etal06}.  For samples of subhalos selected by tidally-truncated mass, objects near the halo center lose a greater percentage of their dark matter mass than objects near the virial radius.  Stellar mass selected samples of satellite galaxies are resistant to this effect since baryonic components are located in the centers of dark matter subhalos and are tightly bound.  The observed radial profile of
satellites in galaxy-sized halos is generally more centrally concentrated than
the subhalo distribution  \citep{chen_etal06} but consistent with the dark matter profile \citep{vandenbosch_etal05,chen_etal06}, although \citet{sales_lambas05} have contrary results.

The color dependence in the spatial distribution of satellite galaxies has been studied extensively in angular distributions.  The angular
distribution of satellites is anisotropic and aligned with the major
axis for red host galaxies and is consistent with isotropic for blue hosts \citep{yang_etal06,agustsson_brainerd08,azzaro_etal07}.  These results are consistent with the picture where the orientation of the major axis of elliptical galaxies is determined by the direction along which subhalos are falling into the galaxy -- along filaments.  The orientation of spiral galaxies is determined by the angular momentum vector, which, in simulations, has shown only poor alignment with the minor axis of the dark matter halo.  A secondary result found in these studies is that red satellites of red hosts show a stronger anisotropy than blue satellites.  This result might be explained by a scenario where satellite color is determined by accretion time;  red satellites were accreted earlier, while blue satellites represent recent infall.  If the major axis of the host galaxy is set early and the orientation of infalling subhalos with respect to the host galaxy changes over time, then objects which accreted early would show a stronger alignment with respect to the host galaxy.  
 
These observational results have been analyzed by comparison to mock galaxy catalogs derived from semi-analytic models  -- simulating galaxies using the mass accretion histories of dark matter halos in a simulation and a variety of assumption about the physics of galaxy formation and evolution.  \citet{agustsson_brainerd08} use mock galaxy catalogs to suggest that the difference in degree of anisotropy between blue and red host galaxies is real and not due to differences in the interloper populations, while \citet{kang_etal07} come to an opposite conclusion using a very different technique --  the color dependence is due to interlopers in the group catalog.  \citet{kang_etal07} also conclude that the difference in alignment between red and blue satellites is due to the masses of the satellites;  red satellites are larger and associated with subhalos which were more massive at the epoch of accretion and which in simulations are accreted more preferentially along the major axis of the halo \citep{libeskind_etal05,wang_etal05}.  The semi-analytic model of \citet{agustsson_brainerd08} suggests that the amount of interloper contamination is much greater for blue satellites than red satellites ($\sim 50\%$ to $\sim15\%$, defining interlopers as objects with a physical separation greater than 500 kpc from the parent galaxy), although the color dependence in the anisotropy of satellites is still observed after subtracting interlopers.  

The color dependence in the radial distribution of satellites has been studied by \citet{sales_etal07} using semi-analytic galaxy catalogs constructed from the Millennium Simulation.  They note that the radial distribution of red satellites is significantly more centrally concentrated than the distribution of blue satellites.  This trend is attributable to accretion time in a scenario similar to one explanation for greater anisotropy in the angular distribution of red satellites compared to blue satellites:  early accreting satellites are stripped of hot gas, stop forming stars, and become redder.  Early accreters are also likely to orbit closer to the center.  

Observationally, the effects of color selection on the radial distribution of satellite galaxies has been previously examined by \citet{sales_lambas04} using data from the Two Degree Field Galaxy Redshift Survey (2dFGRS).  They find a steeper outer slope for satellites of blue parent galaxies than for red primaries.  In addition, red satellites have a distribution that requires a larger core radius than blue satellites and they attribute this result to correlations between the primary and satellite properties.  However, in observational searches for satellites, candidate satellites are chosen based upon their projected separation and velocity difference from the parent galaxy.  Samples of objects chosen in this manner are heavily contaminated by objects which are not satellites -- {\it interlopers}.  In \citet{chen_etal06}, we discussed the importance of interlopers in making relatively unbiased estimates of the projected radial distribution of satellite galaxies and developed a reliable method of interloper subtraction.  Other methods to account for interlopers have been investigated.  For example, \citet{vandenbosch_etal04} exclude interlopers using an iterative, adaptive selection criterion for satellites, and \citet{chen08} model satellites and interlopers together using a halo occupation distribution (HOD) based analytic model for the galaxy correlation function.  The \citet{chen_etal06} approach has the advantage that it may easily be compared to previous results.  

In this paper, we add color selection to the interloper estimation method described in \citet{chen_etal06} and apply it to data from the Sloan Digital Sky Survey (SDSS) spectroscopic sample.  We show that samples of red and blue candidate satellites have different levels of contamination by interlopers, while the model interloper samples for red and blue primaries are similar in color distribution.  We discuss these results in terms of the environmental dependence of satellite galaxies and wider applications to the angular distribution of satellite galaxies and to galaxy formation models.  

\section{Observational Data}

\subsection{The Sloan Digital Sky Survey}

The Sloan Digital Sky Survey (SDSS) includes imaging of the the northern Galactic cap in five bands,
$u,g,r,i,z$, down to $r \sim 22.5$ using a
dedicated 2.5m telescope at Apache Point Observatory in New Mexico in addition to spectroscopic observations for a 
subsample of objects from the imaging catalog \citep{york_etal00,fukugita_etal96,gunn_etal98,hogg_etal01,smith_etal02,strauss_etal02,blanton_etal03b,gunn_etal06,tucker_etal06}.  The SDSS spectroscopy is carried out using optical fibers
positioned in pre-drilled holes on a circular plate, with minimum separation between fibers of
55$\arcsec$, the fiber collision distance. Reobservations of a field can result in observed spectra with separations less than the fiber collision distance, down to the fiber diameter of $3\arcsec$.  The
spectroscopic targets are selected with $r$-band Petrosian magnitudes
$r \leq 17.77$ and $r$-band Petrosian half-light surface brightnesses
$\mu_{50} \leq 24.5$ mag arcsec$^{-2}$.  An automated pipeline measures the
redshifts and classifies the reduced spectra \citep[D. J. Schlegel et
al. 2008, in preparation]{stoughton_etal02,pier_etal03,ivezic_etal04}.\footnote{We use the reductions of the SDSS spectroscopic data performed by D. J. Schlegel et al. (2008, in preparation).
The redshifts found are identical to the redshifts found by an alternative pipeline used for the SDSS Archive Servers (M. SubbaRao et al. 2008, in preparation) over 99\% of the time for Main galaxy sample targets.}

We use the spectroscopic Main galaxy catalog available as Data Release Six \citep[DR6;][]{adelman_mccarthy_etal08}, covering an area of 7425 ${\rm deg^2}$.  Because the SDSS spectroscopy is taken through
circular plates with a finite number of fibers of finite angular size,
the spectroscopic completeness varies across the survey area. The
resulting spectroscopic mask is represented by a combination of disks
and spherical polygons \citep{tegmark_etal04}.  Each polygon also
contains the completeness, a number between 0 and 1 based on the
fraction of targeted galaxies in that region which were observed.  We
apply this mask to the spectroscopy and include only galaxies from
regions where completeness is at least 90\%.  We use $r$-band magnitudes in DR6, built from the NYU Value-Added Galaxy Catalog \citep{blanton_etal05}, normalized to $h$=1, such that $M_{r} =
M_{0.1_{r}} - 5{\rm log}_{10} h$, where $M_{0.1_{r}}$ is the absolute
magnitude K-corrected to $z$=0.1 ({\tt kcorrect v4.1.4}) as described in
\citet{blanton_roweis07}. 

\subsection{Volume-Limited Galaxy Samples}

Following the procedure in \citet{chen_etal06}, we use a volume-limited galaxy sample with a depth of 13,500 km s$^{-1}$, 
corresponding to the limiting redshift of $z=0.045$.   This limit is chosen as a trade-off between 
the volume of the 
sample and the absolute magnitude limit for our satellites, which 
would need to be decreased to brighter magnitudes for more distant 
primaries.  In addition, the limiting magnitude sets a minimum separation at which fiber collisions
become important, which increases with distance. To include more 
distant primaries we would have to sacrifice the ability to probe 
density distributions at small separations.

From our volume-limited sample, we construct a primary sample of isolated host galaxies and a sample of
potential satellites that are projected close to primaries and refer to these two samples as the primary sample and the satellite sample.  Isolated host galaxies are chosen in order to reduce the number of galaxy groups selected and eliminate the contamination from satellites of galaxy group members. Parameters for the criteria follow \citet{prada_etal03} and \citet{chen_etal06} and are listed in Table \ref{tab:select_test}.  The isolation criterion
requires that a primary have only neighbors at least two
    magnitudes fainter within $\Delta R = 0.5 ~h^{-1}$ Mpc and $\Delta V$ = 1000
km s$^{-1}$.\footnote{For this reason, we only search for primaries within the subset of velocities
1000 to 12,500 km s$^{-1}$.}  Potential satellites of any isolated primary must be at least 2 magnitudes fainter than the primary and within $\delta r =0.5 ~h^{-1}$ Mpc and $\delta v = 500{\rm ~km ~s^{-1}}$.  The maximum absolute magnitude for satellites is set by $M_{\rm r,lim} - 5 {\rm
  log} h = 17.77 - DM -K_{0.1}$ in the $r$-band, where the 17.77 is
the flux limit in this band, $DM$ is the distance modulus, and
$K_{0.1}$ is the K-correction at $z$=0.1.  
This gives a limiting absolute
magnitude of $M_r=-17.77$.  The satellites are thus 
limited to the brightest satellite
galaxies, $\sim 0.1L_{*}$.  In order to avoid biasing from a deficit of close pairs of objects, the minimum separation between fibers is 32.9 $h^{-1}$
kpc -- the fiber collision separation at the redshift of the furthest point in our sample.  Finally, we choose galaxies that are in areas that are at least 90\% complete.  For the range $-23 < M_{r} <
-20$, there are 1602 primary galaxies and 690 objects in the satellite
sample with projected radii greater than the minimum separation and
less than 0.5 $h^{-1}$ Mpc.  

\begin{table}
\caption{Selection \& Isolation Criteria}
\label{tab:select_test}
\begin{tabular}{l l}
\hline\hline
Parameters & Value \\
\hline
Constraints on primaries&  ~~$M_{r}<-20$~~\\

Isolation criteria: \\
~~~Magnitude difference & $\Delta M_{r} < 2$ \\
~~~Minimum projected distance, $\Delta R ~(h^{-1}$ Mpc)& 0.5\\
~~~Minimum velocity separation, $\Delta V$ ~(km ${\rm s^{-1}}$)&1000 \\

Satellite sample criteria: \\
~~~Magnitude difference from the primary & $\Delta M_{r} > 2$ \\
~~~Minimum projected distance, $\delta r ~(h^{-1}$ Mpc)&  0.0329\\
~~~Maximum projected distance, $\delta r ~(h^{-1}$ Mpc)&  0.5\\
~~~Maximum velocity separation, $\delta v$ ~(km ${\rm s^{-1}}$)& 500\\

Maximum depth of sample & 13,500 km ${\rm s^{-1}}$ \\

Number of isolated primaries& 1602 \\

Number in satellite sample& 690\\

Limiting magnitude $M_{r}$&$-17.77$\\

Nearby points mock primary criteria: \\
~~~Distance from primary $\Delta R_{\rm corr} ~(h^{-1}$ Mpc) & 1.0-2.0 \\
\hline
\end{tabular}
\end{table}

\subsection{Interloper Subtraction}

There is a fraction of objects in our satellite samples that are not
gravitationally bound to the primaries but are included in the sample
because of projection effects. Throughout this paper, we call such
objects {\it interlopers} (in turn, satellites samples without interloper contamination are referred to as true satellite samples).  Interloper subtraction is discussed in greater detail in \citet{chen_etal06}.  

In semi-analytic galaxy catalogs and in dark matter simulations, interlopers are significant in samples of satellites or DM subhalos.  For example \citet{agustsson_brainerd08} find in a semi-analytic galaxy catalog constructed from the Millennium Simulation -- using a stricter isolation criteria than we use -- that the interloper contamination fraction is $\sim30$\%.  In DM-only simulations, the fraction of interlopers as a function of projected radius rises from a few percent at $R \sim$50 $h^{-1}$kpc to nearly 100\% at  $R =$0.5 $h^{-1}$Mpc \citep[see Fig. 1,][]{chen_etal06}.  

\citet{chen_etal06} developed and tested several methods of subtracting interlopers from the satellite sample statistically.  The projected surface density of candidate satellites, $\langle \Sigma(R) \rangle_{\rm sat}$ is estimated in bins and normalized by the total number of primary galaxies in the sample.  Simple methods of interloper subtraction estimate the corresponding projected surface density in interlopers and subtract it from the interloper contaminated surface density:  $\langle \Sigma(R) \rangle_{\rm int. sub.} = \langle \Sigma(R) \rangle_{\rm sat} -\langle \Sigma(R) \rangle_{\rm int} $.   

A simple and common method of interloper subtraction assumes a uniform distribution of interlopers in space and estimates the interloper surface density by counting the number of objects that satisfy the satellite criteria but are located around empty points in the sky, or mock primary galaxy positions.  The mock satellite samples created from these points constitute samples of pure interlopers:  there are no satellites in these samples.  Hereafter, we generically refer to these estimates of purely interloper -- mock satellite -- samples as model interlopers.  We then fit the surface density profile by a power-law,
\begin{equation}
\Sigma (R) = A R^{\alpha},
\end{equation}
with slope, $\alpha$, and normalization, $A$.  In DM simulations, the difference in slope between the satellite sample with no interloper subtraction and the true satellite sample is $\Delta \alpha \sim 0.5$.  The difference between the satellite sample and the interloper-subtracted sample using this simple estimate is $\Delta \alpha \sim 0.1$, a vast underestimate of the fraction of interlopers.  This simple interloper subtraction method fails because it oversamples voids compared to clustered areas -- there are a lot more isolated, empty points in voids than elsewhere.  Most galaxies, however, reside in clustered areas, even for our sample of isolated galaxies.  

To account for this clustering effect, our preferred method -- the `nearby points' method -- samples the environments of our primary galaxies.  We pick mock primary galaxy positions 
that are between $1 ~h^{-1}$~Mpc and $2~h^{-1}$~Mpc from a real primary:  the minimum separation avoids sampling real satellites.  All mock primary points satisfy the same isolation
criteria as our sample of primaries, and the size of this mock primary sample
is 20 times the number of primary galaxies.\footnote{In our data there is a total of  3325 model interlopers.}

Model interlopers from the area near actual primaries better approximate the interloper contamination in the satellite sample.  For example, mock galaxy catalogs show that the sample of true interlopers contains objects with velocities correlated to the primary:  the distribution of relative velocities between interloper and primary is not constant \citep[see  Fig. 3,][]{chen_etal06}.  This structure in relative velocities of true interlopers is reproduced in the relative velocities of model interlopers with respect to primaries using the nearby points method and cannot be reproduced using the simple interloper method outlined above.  In previous tests on mock galaxy catalogs, the nearby points method returns a best-fit power-law slope within $\sim0.1$ of the true satellite distribution.  In \citet{chen_etal06}, for a set of SDSS galaxies using the same isolation and satellite criteria as used in this paper, the nearby points  method returns a best-fit power-law slope of $\alpha = -1.58 \pm 0.11$.  The errors quoted are statistical and do not include the $\Delta\alpha \sim 0.1$ bias inherent to the method.  Accounting for the systematic error would give $\alpha \approx -1.7 \pm 0.1$.  Using our larger DR6 data set, we calculate the surface density of objects in six bins of width 0.067 $h^{-1}$ Mpc, and we find a consistent result for the nearby points method, $\alpha = -1.52 \pm 0.07$ (see Table \ref{tab:slope_all}).  These results are also consistent with \citet{sales_etal07} who estimate the radial distribution of satellite galaxies using semi-analytic galaxy catalog constructed from the Millennium Simulation and find $\alpha = 1.55 \pm 0.08$.  

As a final note, we might expect that the interloper-subtracted surface density profile would be best fit by the \citet[][NFW]{navarro_etal97} profile that is used to describe the distribution of matter in dark matter halos.  However, our minimum separation is on the order of the expected scale radius for our parent galaxy halos, making fitting for the scale radius impractical.  

\section{Results} 

\begin{figure}
\centering
\resizebox{3in}{!}
	{\includegraphics{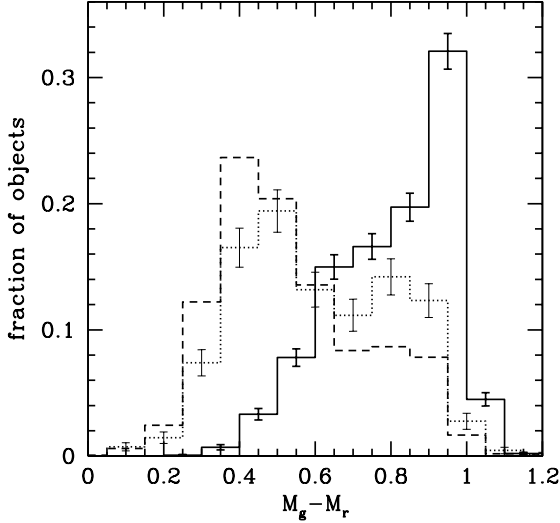}}
\caption{The color distribution for the primary sample (solid), satellite sample (dotted), and model interloper sample (dashed).  Red satellites are defined to have $M_g-M_r > 0.65$, while red primaries have colors $M_g-M_r > 0.9$.  Poisson errors are shown for the primary and satellite sample.  The corresponding Poisson errors for the model interloper sample are significantly smaller than for the satellite sample.  }
\label{fig:color_hist}
\end{figure}

The $M_g-M_r$ color distributions for the primary, satellite, and model interloper samples are shown in Fig. \ref{fig:color_hist}.  The primary sample is redder than the satellite sample with a large red peak and a bluer tail, while the satellite sample has a vaguely bimodal distribution, split at $M_g-M_r \sim 0.65$.  The model interloper distribution appears tilted to bluer objects compared with the distribution of the satellite sample.  

\subsection{Red and Blue Satellites}

The differences in color distribution between the satellite sample and the model interloper sample suggests that a greater fraction of blue objects in the satellite sample will be interlopers than of red objects.  Defining red satellites to $M_g-M_r > 0.65$, of the 690 objects in the satellite sample, 285 are red and 405 are blue.  By comparison to the model interloper sample, the fraction of the objects that are interlopers in the whole sample is 24\%, while the percentages for the red and blue satellite sample are 16\% and 30\%, respectively.  

\begin{figure}
\centering
\resizebox{3in}{!}
	{\includegraphics{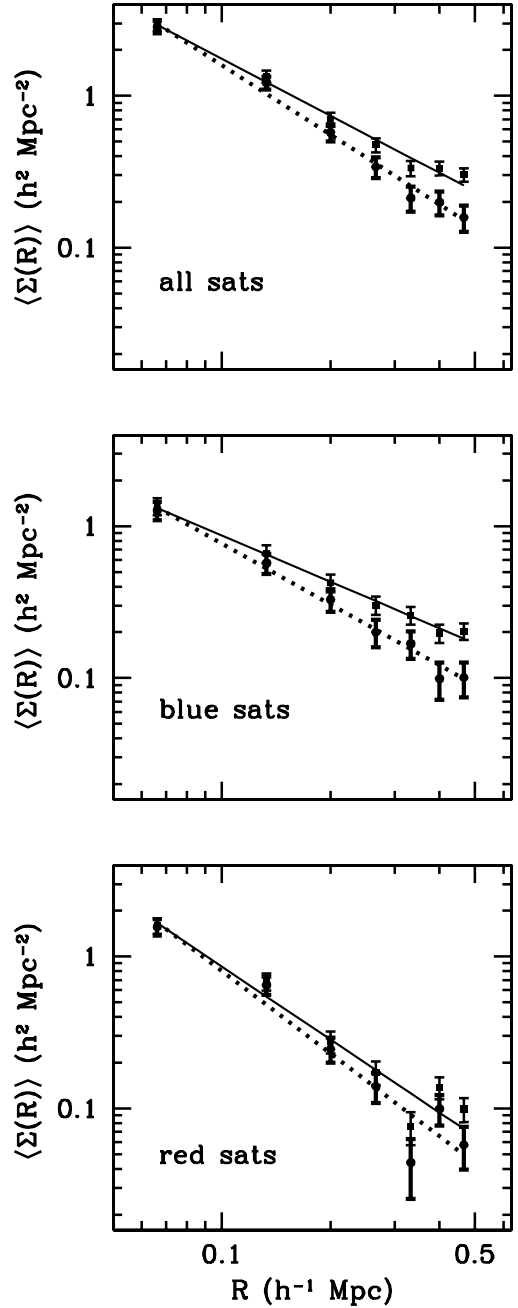}}
\caption{The surface density of the satellite sample (squares) and the interloper-subtracted satellite sample (circles) for all satellites {\it (top)}, blue satellites {\it (center)}, and red satellites {\it (bottom)}.  The best fit power-laws are also plotted with solid lines for the satellite sample and dotted lines for the interloper-subtracted sample. }
\label{fig:int_sub}
\end{figure}

This difference affects the estimates of the slope of the power-law fit to the radial distribution.  In Figure \ref{fig:int_sub}, the biasing of the best-fit power-law is shown for all satellites and blue and red satellites.  The distribution of blue satellites is significantly flatter than the distribution of red satellites, and the distribution of blue satellites is more effected by interlopers than the distribution of red satellites.  

\begin{figure}
\centering
\resizebox{3in}{!}
	{\includegraphics{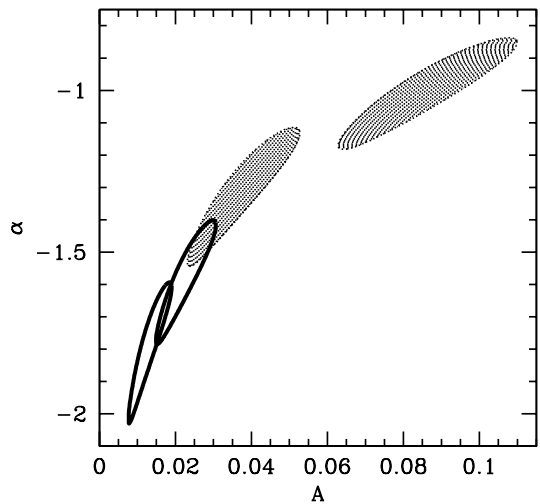}}
\caption{The 68\% confidence intervals for the normalization ($A$) and slope ($\alpha$) of the power-law, $\Sigma (R) = A R^{\alpha}$.  Shaded contours are for blue satellites;  solid line contours are for red satellites.   The contour for interloper-subtracted samples are in each case at steeper slopes than their satellite sample counterparts.}
\label{fig:chisq_plot}
\end{figure}

In the interloper-contaminated satellite samples, the best-fit slopes for blue and red satellites (shown in Table 2 and Figure \ref{fig:chisq_plot}) are inconsistent, with a much shallower slope for blue satellites.  After interloper subtraction, although the difference in slopes is smaller, the blue slope is still significantly shallower than the red slope, and the slopes are inconsistent given both the marginalized errors and their $1\sigma$ confidence intervals (see Fig. \ref{fig:chisq_plot}).  The projected radial distribution of red satellites is as steep as might be expected for the dark matter density distribution in halos which host galaxies such as are found in our primary sample.  The distribution of blue satellites, on the other hand, is as shallow as might be expected for the subhalo distribution in the same host halos \citep[see, for comparison,][]{chen_etal06}.  

\begin{table}[h]
\caption{Best-Fit Power-Law Slopes to the Surface Density Profile of Satellites}
\label{tab:slope_all}
\begin{tabular}{lcc}
\hline\hline
Input Data & ~~~Satellite Sample~~~& ~~~Interloper Subtracted~~~\\
\hline
\noalign{\smallskip}
all &  $-1.24 \pm 0.06$   & $-1.52 \pm 0.07$  \\
\noalign{\smallskip}
blue satellites& $-1.01 \pm 0.08$  & $-1.32 \pm 0.10$ \\
~~~~~bright & ... & $-1.42 \pm  0.15$ \\
~~~~~faint & ... &  $-1.25 \pm 0.13$ \\
\noalign{\smallskip}
red satellites&  $-1.59 \pm 0.09$    & $-1.80 \pm 0.10$ \\
~~~~~bright & ... & $-1.78 \pm 0.14$\\
~~~~~faint & ... & $-1.86 \pm 0.15$\\
\noalign{\smallskip}
blue primaries& $-1.18 \pm 0.10$  & $-1.54 \pm 0.12$ \\
~~~~~bright & ... & $-1.67 \pm 0.15$\\
~~~~~faint & ... & $-1.35 \pm 0.20$\\
\noalign{\smallskip}
red primaries&  $-1.27^{+ 0.07}_{-0.08}$    & $-1.50\pm 0.09$ \\
~~~~~bright & ... & $-1.34 \pm 0.11$\\
~~~~~faint & ... & $-1.82^{+0.15}_{-0.14}$\\
\noalign{\smallskip}
\hline
\end{tabular}
\end{table}

\begin{figure}
\centering
\resizebox{3in}{!}
	{\includegraphics{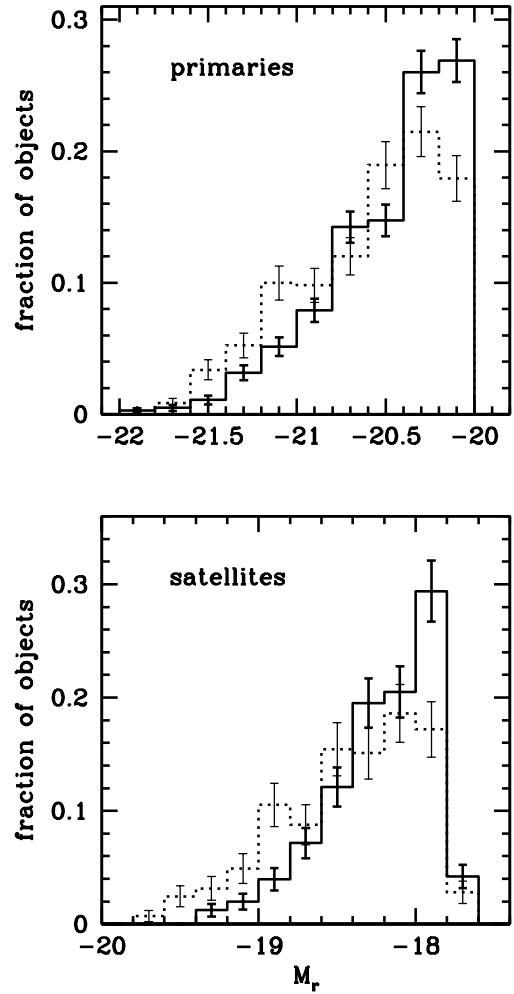}}
\caption{The luminosity distribution with corresponding Poisson errors of primaries ({\it top}) and objects in the satellite sample ({\it bottom}), where red objects and blue objects are shown separately, in dotted and solid lines respectively.  Bright and faint satellites are split at $M_{r} = -18.3$, while bright and faint primaries are split at $M_{r} = -20.8$. }
\label{fig:test6}
\end{figure}

In general, we expect blue satellites to be fainter than red satellites, so it is useful to attempt to disentangle the effects of luminosity from those of color.  While \citet{chen_etal06} tests luminosity dependence (which is further discussed in Section \ref{sec:conclusions}), they do not have sufficient numbers of objects to test color and luminosity together.  Figure \ref{fig:test6} shows the luminosity distributions for the blue and red satellites and primaries.  We split each color sample of satellites into a faint and a bright sample at $M_{r} = -18.3$.  Table \ref{tab:slope_all} shows that the interloper-subtracted surface density profiles of faint and bright samples are consistent for color-selected samples, i.e., color is the dominant attribute in the radial distribution of satellites.  

\subsection{Red and Blue Primaries}

\begin{figure}
\centering
\resizebox{3in}{!}
	{\includegraphics{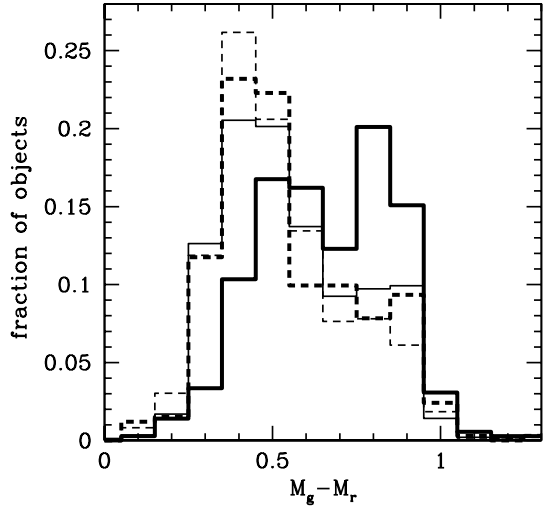}}
\caption{The color distribution for the satellite sample and model interloper sample for red primaries (thick and thin solid lines, respectively) and the color distribution for the satellite sample and model interloper sample for blue primaries (thick and thin dashed lines, respectively). }
\label{fig:color_hist_pri}
\end{figure}

\begin{figure}
\centering
\resizebox{3in}{!}
	{\includegraphics{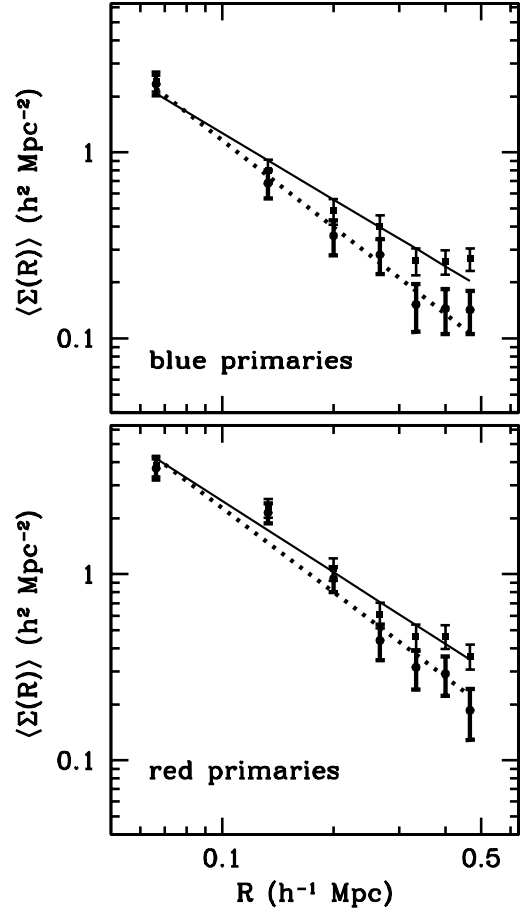}}
\caption{The surface density of the satellite sample (squares) and the interloper-subtracted satellite sample (circles) for satellites of blue primaries {\it (top)} and of red primaries {\it (bottom)}.  The best fit power-laws are also plotted with solid lines for the satellite sample and dotted lines for the interloper-subtracted sample.  }
\label{fig:int_sub_pri}
\end{figure}

We split the 1602 primary galaxies by color at $M_g-M_r =0.9$.  The 591 red primary galaxies have 358 objects in their satellite sample, while the 1011 blue primary galaxies have 332 candidate satellites.  The interloper percentages for the red and blue primary sample are 21\% and 28\%, respectively.  Red primaries of $M_g-M_r > 0.9$ have a larger fraction of red objects in their satellite sample than than blue primaries (52\% to 30\%) and, subsequently, more true satellites (see Fig. \ref{fig:color_hist_pri}).  This can also be seen in Fig. \ref{fig:int_sub_pri}, where the amplitude of the satellite profile of red primaries is greater than that of the blue primary profile.  After interloper subtraction, the slopes of the density profiles of satellites for blue and red primaries are consistent.   

Figure \ref{fig:test6} additionally shows that red primaries are in general brighter than blue primaries.  Splitting the color-selected primaries into faint and bright samples at $M_{r} = -20.8$, Table \ref{tab:slope_all} shows significant differences for the distribution of satellites of faint and bright primaries, depending on the color of the primary.  Red, brighter primaries have an interloper-subtracted satellite profile that is shallower than the distribution of satellites in red, fainter primaries, $\alpha = -1.34\pm 0.11$ to $\alpha = -1.82^{+0.15}_{-0.14}$.   This is not unexpected; from numerical simulations we expect the satellite distribution to scale with the mass distribution of the primary, and larger primaries have smaller concentrations leading to shallower slopes at the radii at which we measure.  Interestingly, blue primaries show the opposite dependence:  the distribution of satellites around brighter primaries seems  to have a steeper slope than around fainter primaries.  This trend is accompanied by a change in the fraction of red satellites:  38\% of the satellite sample for bright, blue primaries are red, while 23\% of the corresponding sample for faint, blue primaries are red.  The errors, however, are nearly as large as the discrepancy between the samples, and larger samples will be required to confirm this result.  In addition, the slope for satellites of bright, blue primaries is steeper than that for bright, red primaries, which may reflect color dependence in using luminosity as a proxy for mass.  

Similar results can be seen if we split both the primary and the satellite samples by color.  In Table \ref{tab:pri_sat}, we show that red satellites around blue primaries have a larger interloper fraction and a steeper power-law slope than found for red satellites of red primaries.  The difference in power-law slope can be attributed to the same luminosity dependence discussed previously:  red primaries are more massive and have mass and satellite distributions described by smaller concentrations.  Interloper contamination increases from small separations to large, and for larger primaries, the range of radii we probe preferentially sample areas with smaller level of interloper contamination.  On the other hand, for blue satellites of red and blue primaries, we find very similar interloper fractions and shallow power-law slopes.  Unfortunately, in all cases, statistics are poor and future larger samples will be required for statistical significance to be ascribed.  

\begin{table}[h]
\caption{Primaries and Satellites Selected by Color}
\label{tab:pri_sat}
\begin{tabular}{lccc}
\hline\hline 
Input Data  & Number in & Interloper & Power-Law \\
(Primary-Satellite) & Satellite Sample  & Fraction & Slope \\
\hline
\noalign{\smallskip}
blue  - blue & 232  & 0.30 &   $-1.33 \pm 0.15$     \\
blue  - red &  100 & 0.22 &$-1.98^{+0.19}_{-0.20}$\\
red  - blue & 173 & 0.30 & $-1.26\pm 0.14$  \\
red - red & 185 & 0.12   &$-1.71 \pm 0.12$  \\
\noalign{\smallskip}
\hline
\end{tabular}
\end{table}

\subsection{Environmental Dependence}

Despite the difference in satellite samples, the color distribution of the model interloper samples of red and blue primaries are similar to each other and to the color distribution of the satellite sample of blue primaries (see Fig. \ref{fig:color_hist_pri}).  Restated, although red primaries are found in more clustered environments, the environments of red and blue primaries are not noticeably different as measured by the color distribution of model interlopers.  

\begin{figure}
\centering
\resizebox{3in}{!}
	{\includegraphics{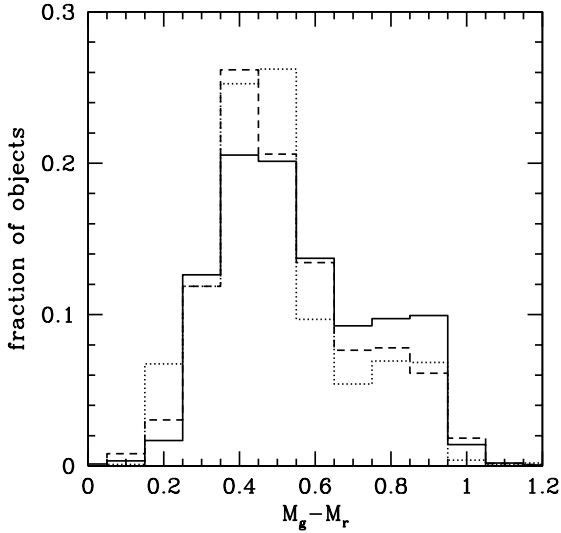}}
\caption{The color distribution for the model interloper samples for red primaries (solid), blue primaries (dashed), and for sampling isolated points greater than 2$h^{-1}$ Mpc from primaries (dotted).  The average surface number density of objects in these samples is 0.159, 0.117, and 0.0415 $h^2 ~{\rm Mpc}^{-2}$. }
\label{fig:color_hist_ints}
\end{figure}

Red primaries live in more clustered environments than blue primaries as measured in the average surface number density of model interlopers:  0.159 $h^2 ~{\rm Mpc}^{-2}$ for red primaries and 0.117 $h^2 ~{\rm Mpc}^{-2}$ for blue primaries.  In the previous section, we noted that these slightly different environments have similar color distributions of faint objects, even though we do not use fixed luminosity criteria for these objects.  We test one less clustered environment, sampling isolated points that lie outside of a 2 $h^{-1}$ Mpc radius from our isolated galaxies (outside the criteria for our interloper subtraction method).  Here, the average surface density is 0.042 $h^2 ~{\rm Mpc}^{-2}$.  All three model interloper distributions are plotted in Fig. \ref{fig:color_hist_ints}, in which all environments have similar color dependences with a possible trend to bluer objects in less dense fields.  

\section{Conclusions}
\label{sec:conclusions}

In \citet{chen_etal06}, we constrained the projected radial distribution for isolated galaxies and found that their power-law slopes are steeper than the expected slopes for the distribution of dark matter subhalos and may be as steep as the density profile of the host dark matter halos.  We reproduce this result with a survey area that is $\sim 50\%$ larger.  However, the distribution of satellite galaxies shows some significant dependence on color.  

When samples of candidate satellites are split by color, we see that blue objects are more likely to be interlopers than red objects.  The observed estimated interloper contamination of red objects in the satellite sample is 16\%, while that of blue objects is 30\%.  \citet{agustsson_brainerd08} produce a difference of 15\% and 50\%, respectively, in the interloper contamination of red and blue satellites samples, using a semi-analytic galaxy formation model.  It is, then, not appropriate to assume that red and blue satellites have the same level of interloper contamination and this must be taken into account in testing the color dependence in the angular distribution of satellite galaxies.  This result also suggests that the prevalence of faint red galaxies could be developed into a method to find small groups of galaxies in a manner similar to the method by which the red sequence of early-type galaxies is used to find galaxy clusters \citep[e.g.,][]{gladders_yee00}.  

Both with and without interloper subtraction, the radial profile of blue satellites is significantly shallower than that of red satellites.  The best-fit power-law slope of interloper-subtracted blue satellites is $\alpha =1.34 \pm 0.10$ and the that of red satellites is $\alpha = -1.80 \pm 0.10$.  This result is consistent with the trend found by \citet{sales_etal07} who test color dependence in semi-analytic galaxy catalogs constructed from the Millennium Simulation, finding a more centrally concentrated radial distribution of red satellites than of blue satellites.  This trend is repeated when they select by a proxy for accretion time -- whether the satellite retains a DM halo or whether it has been tidally destroyed.  

Blue satellites are generally fainter than red satellites.  Correspondingly, \citet{chen_etal06} found that the best-fit power-law slope for bright satellites is steeper than that of faint satellites (cut at $M_r = -18.28$) -- although without statistical significance -- in a volume-limited sample.  In a flux-limited sample, the reverse relation was found;  however, this sample used a luminosity cut as faint as the faintest satellites in our sample, at $M_{r} = -17.76$, and so is not directly comparable.  When we cut the satellites samples by color and luminosity, we find that the dominant effect is from color;  the best-fit power-law slopes of bright and faint samples of red satellites are consistent as are those of bright and faint blue satellites.  

The power-law slope for red satellites is as steep as might be expected for the dark matter density distribution of halos which host galaxies like those found in our primary sample.  On the other hand, the power-law slope for blue satellites is as shallow as the expected subhalo distribution.  The shallowness of the subhalo profile is attributed to tidal stripping. This is unlikely to effect satellite galaxies, since they are located at the centers of the dark matter subhalos.  The shallower profile for blue satellites as compared to red satellites, then, might be interpreted as consistent with the scenario where satellite color is determined by accretion time:  red satellites were accreted earlier, while blue satellites represent more recent infall.  Satellites are expected to be redder in the inner regions of parent halos due to environmental processes that shut off star formation (ram pressure stripping, strangulation, etc.).  This morphological segregation has also been observed in more massive clusters and in the fainter satellites found in the Local Group.  

When splitting the satellite sample by primary color, red primaries have a significantly larger fraction of red satellites and somewhat smaller interloper fractions than blue primaries.  After interloper subtraction, the best-fit power-law slopes of satellites of red and blue primaries are consistent within errors.  Red primaries are, on average, more luminous than blue galaxies.  Correspondingly, \citet{chen_etal06} found that the best-fit power-law slope for satellites bright primaries is consistent with that of satellites of faint primaries (cut at $M_r = -21$).  When we cut the primary sample by luminosity and color, the trend with luminosity is different for red and blue primaries.  The slope of satellites of bright, red primaries is shallower than that of faint, red primaries, a relation probably dominated by the mass of the parent halo as brighter primary galaxies reside in bigger parent halos which have mass distributions characterized by smaller concentrations.  The slope of satellites of bright, blue primaries is steeper than that of faint, blue primaries, as the fraction of red satellites drops with primary luminosity.  

The color distribution of objects in model interloper samples are similar, regardless of their environment (as measured by average surface density).  While robust conclusions cannot be drawn as to how similar these distributions are, it suggests that there are fundamental differences between satellite galaxies and faint galaxies in the field.  Intriguingly, this blue-tilted color distribution also resembles that of the satellites sample of blue primaries.  Better understanding of the processes that effect the color of faint objects and the radial distribution of satellite galaxies will require further studies.  


\begin{acknowledgements}
We would like to thank Andrey Kravtsov, Francisco Prada, Michael Blanton, and Erin Sheldon 
for their suggestions and invaluable contributions to understanding interloper contamination and systematics in the data.

In addition, we would like to acknowledge the anonymous referee for many helpful suggestions.  

Funding for the Sloan Digital Sky Survey (SDSS) has been provided by
the Alfred P. Sloan Foundation, the Participating Institutions, the
National Aeronautics and Space Administration, the National Science
Foundation, the U.S. Department of Energy, the Japanese
Monbukagakusho, and the Max Planck Society. The SDSS Web site is
http://www.sdss.org/.

The SDSS is managed by the Astrophysical Research Consortium (ARC) for
the Participating Institutions. The Participating Institutions are The
University of Chicago, Fermilab, the Institute for Advanced Study, the
Japan Participation Group, The Johns Hopkins University, Los Alamos
National Laboratory, the Max-Planck-Institute for Astronomy (MPIA),
the Max-Planck-Institute for Astrophysics (MPA), New Mexico State
University, University of Pittsburgh, Princeton University, the United
States Naval Observatory, and the University of Washington.
\end{acknowledgements}

\bibliographystyle{aa}
\bibliography{research}

\end{document}